\documentclass[12pt]{iopart}
\usepackage{epsf,amssymb}
\def\bfone{\relax{\rm 1\kern-.35em 1}}
\begin{document}
\begin{flushright}
  KUL-TF-2002/13\\
ITP-UU-03/01\\
SPIN-03/01\\
\end{flushright}
\vspace{.5cm}
\title[]{$\mathcal{N}=2$ supergravity models with stable de Sitter vacua}

\author{Pietro Fr{\'e}\dag\ , \underline{Mario Trigiante}\ddag \ and Antoine Van Proeyen{}$^*$
\footnote[3]{Work supported in part by the European Community Marie Curie
Fellowship under contract  HPMF-CT-2001-01276.} }

\address{\dag\
 Dipartimento di Fisica Teorica, Universit{\`a} di Torino, $\&$ INFN - Sezione di Torino\\
via P. Giuria 1, I-10125 Torino, Italy}

\address{\ddag\ Spinoza Institute, Leuvenlaan 4, NL-3508 Utrecht, The
Netherlands}

\address{${}^*$\ Instituut voor theoretische fysica, Katholieke Universiteit Leuven, \\ B-3001 Leuven, Belgium}

\begin{abstract}
In the present talk I shall review the construction
of $\mathcal{N}=2$ supergravity models exhibiting stable de Sitter vacua.
These solutions represent the first instance of stable backgrounds with positive cosmological constant in the framework of extended supergravities ($\mathcal{N}\ge 2$). After briefly reviewing the role of de Sitter space--times in inflationary cosmology, I shall describe the main ingredients which were necessary for the construction of gauged $\mathcal{N}=2$ supergravity models admitting stable solutions of this kind.
\end{abstract}



\maketitle

\section{Introduction}
De Sitter space--time $dS$ can be characterized as the maximally
symmetric solution of Einstein's equations with positive cosmological
constant $\Lambda$. $D$-dimensional de Sitter space can be represented as
the hyperboloid in $\mathbb{R}^{1,D}$ defined by the following equation:
\begin{eqnarray}
(X^0)^2-\sum_{i=1}^D\,(X^i)^2 &=& -H^{-2}\,.
\end{eqnarray}
The above geometry in $D=4$ is a solution of Einstein's equation
in the vacuum with a cosmological constant $\Lambda=3H^2>0$ ($8\pi
G_N=1$) or alternatively can be interpreted as the geometry
generated by the interaction with gravity of a system with uniform
energy density $\rho=3H^2$ and pressure $p=-\rho$.\par One of the
interesting properties of de Sitter space--time is to admit
locally a description in terms of a
FRW metric which in $D=4$ has the following form:
\begin{eqnarray}
ds^2 &=& dt^2-a(t)^2\left(\frac{dr^2}{1-k
r^2}+r^2d\theta^2+r^2\sin{(\theta)}^2 d\varphi^2\right).
\end{eqnarray}
Since the scale factor $a(t)>0$ fulfils the equation
$\ddot{a}/a=\Lambda/3=H^2
>0$, the above metric describes  the evolution of a spatially closed
$k=1$, flat $k=0$ or open $k=-1$ universe which undergoes an
accelerated expansion.\par The recent interest attracted by de Sitter space--time
 in the field of cosmology stems from the present
experimental evidence \cite{experiment} that our universe is expanding at
an accelerated rate and thus can be consistently described as either
being in a  de Sitter regime with $\Lambda\sim 10^{-120} M_p^4$ or
evolving towards one \cite{kallosh:linde2}. The recent experimental
observations moreover seem to indicate that our universe is spatially
flat $\Omega=1$ and that the dominating fraction ($0.7$) of its energy
has the form of a {\it dark energy} whose equation of state is close to
$w=p/\rho =-1$. These results corroborate inflationary scenarios (see
\cite{linde90} for a review) according to which our universe underwent
stages of accelerated expansion triggered by the slow evolution of a
uniform scalar field $\phi$ whose energy is dominated by a positive
potential $V(\phi)>0$, $V(\phi)\gg(\dot{\phi})^2$ and should account for
the predicted dark energy. These models may have de Sitter critical
points $\phi_0$ in which $\partial_\phi V_{|\phi_0}=0$ and the value of
the potential is non vanishing $\Lambda=V(\phi_0)>0$ so as to provide the
effective cosmological constant for the corresponding space--time.\par
Recently, considerable effort has been devoted towards deriving
inflationary models from fundamental theories as string/M--theory. A
field theory framework which is most closely related to string/M--theory
is provided by extended supergravities ($\mathcal{N}\ge 2$) and therefore
it is natural as a first step to consider the possible cosmological
models which can be constructed from these latter theories. The high
degree of supersymmetry, which characterizes extended supergravities,
implies that the only way of deforming the action in order to allow for a
non trivial scalar potential is to promote some of the global symmetries
to local symmetries of the theory ({\it gauging}). There are several
instances in the literature of $\mathcal{N}=2,\,3,\,4,\,8$ models
admitting de Sitter solutions \cite{dSexamples}, \cite{deRoo:Wagemans}
and most of them are constructed by means of a non--compact
gauging.\footnote{See \cite{HulletAl} for an interpretation of these
non--compact local symmetries in the $\mathcal{N}=8$ theory as the
isometry group of a class of non--compact internal manifolds on which
string/M--theory is dimensionally reduced to four dimensions.} These
examples have some common features:
\begin{itemize}
\item{the Casimir $C_2$ of the de Sitter isometry group ${\rm SO}(1,4)$, which measures the scalar mass spectrum in units of $\Lambda$  \cite{kallosh}, has an integer spectrum on the scalar excitations:
\begin{eqnarray}
\langle C_2\rangle = \frac{m^2}{H^2}=\frac{3\, m^2}{\Lambda}=k\in
\mathbb{Z}\,. \label{casimir}
\end{eqnarray}
}
\item{there was always a tachyon with $k=-6$ and therefore all these solutions were unstable \cite{kallosh:linde1}.}
\end{itemize}
As shown in \cite{our} in the context of $\mathcal{N}=2$ supergravity the latter is not a general property since a class of models exists which admit stable de Sitter vacua.
\section{$\mathcal{N}=2$ supergravity and gauging}

\begin{figure}
\begin{center}
\epsfxsize =10cm
%\vskip 10cm
{\epsffile{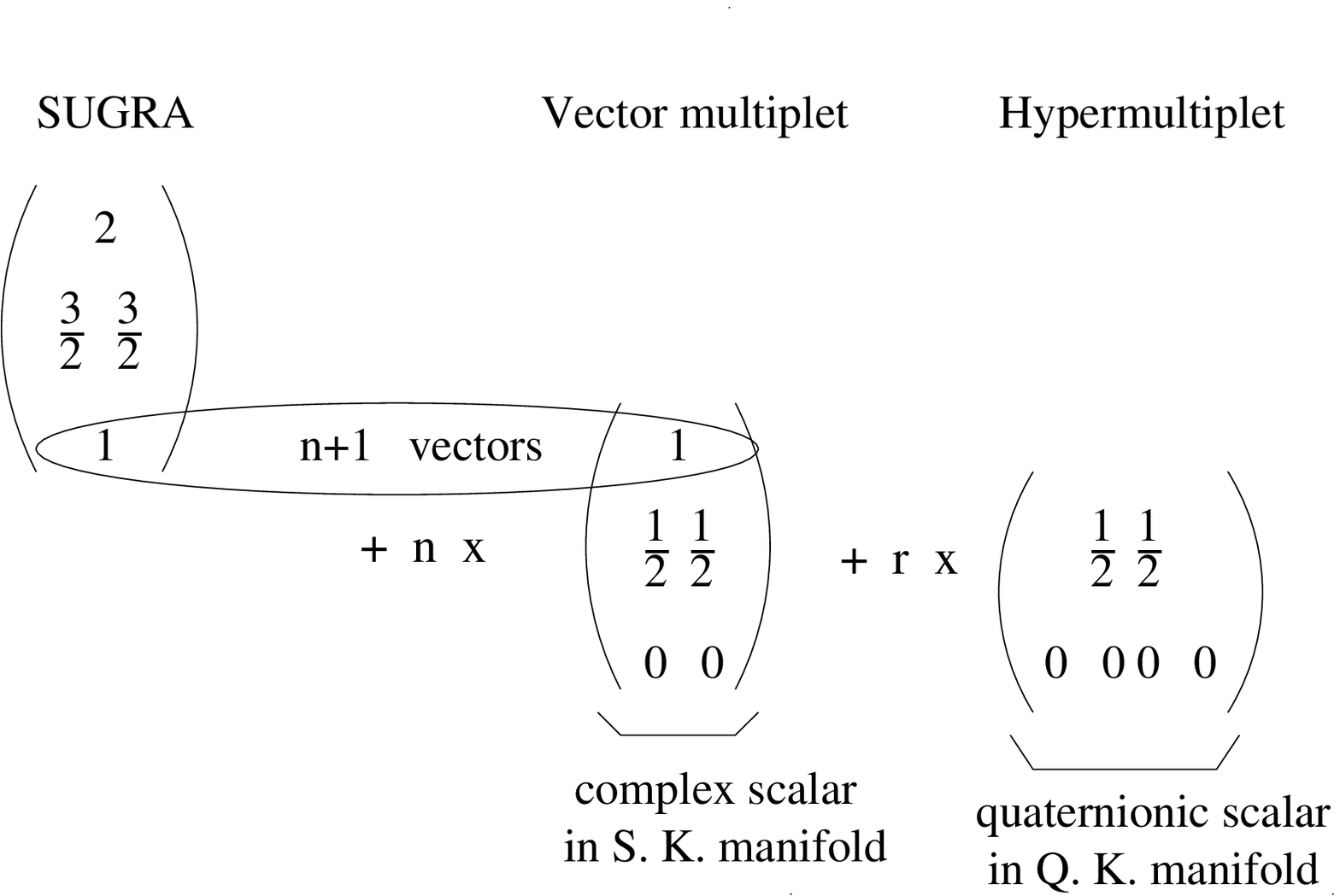}
%\vskip -8.5cm
}
% \hskip 2cm
%  \unitlength=1.1mm
\caption{\label{N2}}
\end{center}
\end{figure}

The general $\mathcal{N}=2$ supergravity of \cite{deWit:Lau:VP,n2gen}
describes a gravitational multiplet coupled to $n$ vector multiplets and
$r$ hyper--multiplets as illustrated in figure \ref{N2}.
The vector multiplets contain $n$ complex scalar fields $z^\alpha$ which
span a Special K{\"a}hler manifold $\mathcal{M}_{SK}$ of complex dimension
$n$ while the $4r$ real scalars $q^u$ contained in the hyper--multiplets
are coordinates of a Quaternionic--K{\"a}hler manifold $\mathcal{M}_{QK}$ of
quaternionic dimension $r$. The model describes $n+1$ vector fields. The
scalar manifold of the theory is
$\mathcal{M}_{scal}=\mathcal{M}_{SK}\times \mathcal{M}_{QK}$. Its
isometry group  $\mathbb{G}={\rm Isom}(\mathcal{M}_{scal})$ is promoted
to global symmetry group of the field equations and Bianchi identities
once its action is defined as a simultaneous transformation on the scalar
fields  and on the vector of electric $F^\Lambda_{\mu\nu}$ and magnetic
$G_{\Lambda|\mu\nu}$ field strengths. This latter is a generalized
electro--magnetic duality defined by the embedding of $\mathbb{G}$ inside
${\rm Sp}(2n+2,\mathbb{R})$: {\small
\begin{eqnarray}
\mathbb{G}&\rightarrow &\cases{\mbox{non--linear action on
$\phi^I=\{z^\alpha,\,q^u\}$}\,,\begin{array}{l}
  \alpha =1,\ldots ,n\,, \\
  u=1,\ldots ,4r\,,
\end{array}  \cr
\mbox{linear action } \left(\matrix{F^\Lambda_{\mu\nu}\cr G_{\Sigma|\mu\nu}}\right)\rightarrow {\bf g}\cdot  \left(\matrix{F^\Lambda_{\mu\nu}\cr G_{\Sigma|\mu\nu}}\right)\,,\,\,\Lambda,\Sigma=0,\dots, n\,,}\nonumber\\&&\nonumber\\
{\bf g} &=& \left(\matrix{A & B\cr C & D}\right)\in
\mathbb{G}\hookrightarrow {\rm Sp}(2(n+1),\mathbb{R})\,.
\end{eqnarray}}
The geometry of $\mathcal{M}_{SK}$ is locally described by the choice of the coordinates $\{z^\alpha\}$ and of a $2(n+1)$--dimensional section $V(z^\alpha)$
of a holomorphic symplectic bundle on $\mathcal{M}_{SK}$, which fixes the coupling between the scalars and the vector field strengths in the action. In terms of  $V(z^\alpha)$ the K{\"a}hler potential has the following  form:
{\small \begin{eqnarray}
 \mathcal{K}(z,\overline{z})&=&-{\rm log}\left({\rm i}\langle V\vert \overline{V}\rangle\right) \,\,\,;\,\,\,\,
V(z^\alpha)= \left(\matrix{X^\Lambda(z^\alpha)\cr
F_{\Sigma}(z^\alpha)}\right) \nonumber\,\,;\,\, \langle V\vert
\overline{V}\rangle=\overline{X}^\Lambda F_{\Lambda}-X^\Lambda
\overline{F}_{\Lambda}\,.
\end{eqnarray}}
In the ungauged version of the theory,
namely when no field is charged with respect to local transformations gauged
by vector fields, no scalar potential is allowed by supersymmetry.
The only way to produce a non trivial scalar potential and therefore non
trivial vacua is to promote a suitable subgroup $\mathcal{G}$ of $\mathbb{G}$
which is also global symmetry of the action, to local symmetry. This gauging
procedure roughly consists of the following steps:
\begin{itemize}
\item{There have to exist vector fields transforming under $\mathcal{G}$ in the co--adjoint representation. They will become gauge vectors associated with the generators of $\mathcal{G}$}\item{Minimal couplings are introduced by substituting ordinary derivatives with covariant ones. A different coupling constant
$e_k$
 is associated with each simple factor in $\mathcal{G}$.}
\item{Supersymmetry invariance of the theory requires the addition to the action of fermion mass terms and of a scalar potential $V(\phi)\neq 0$ while the fermion supersymmetry transformations have to be modified by adding suitable shifts (fermion shifts).}
\end{itemize} Duality invariance is broken by the gauging procedure since
the minimal couplings involve only the elementary vector fields and not their duals. \par
 The action of the gauge group $\mathcal{G}$ on $\mathcal{M}_{QK}$ is described in terms of Killing vectors
$k_\Lambda=k_\Lambda^u\partial _u$ fulfilling the following commutation
relation: $\left[k_\Lambda,k_\Sigma\right]=-f_{\Lambda\Sigma}{}^\Delta
k_\Delta$, $f_{\Lambda\Sigma}{}^\Delta$ being the  structure constants of
$\mathcal{G}$. Each vector $k_\Lambda^u$ can be expressed in terms of
{\it tri--holomorphic} moment maps $\mathcal{P}^x_\Lambda$, where
$x=1,2,3$ run over the adjoint representation of the SU$(2)$--holonomy
group of $\mathcal{M}_{QK}$. The   following condition must hold:
\begin{eqnarray}
2\, k^u_\Lambda\, k^v_\Sigma\, R^x_{uv}-\epsilon^{xyz}\,
\mathcal{P}^y_\Lambda\, \mathcal{P}^z_\Sigma &=&
f_{\Lambda\Sigma}{}^\Delta \,\mathcal{P}^x_\Delta\,, \label{equivariance}
\end{eqnarray}
where $ R^x_{uv}$ is the curvature of the SU$(2)$ bundle over
$\mathcal{M}_{QK}$. In absence of hyper-multiplets ($r=0$), $ R^x_{uv}$
is trivially zero and we can have a solution of eq.~(\ref{equivariance})
with $\mathcal{P}^x_\Lambda\equiv \mbox{const}$:
\begin{eqnarray}\epsilon^{xyz} \mathcal{P}^y_\Lambda \mathcal{P}^z_\Sigma &=&- f_{\Lambda\Sigma}{}^\Delta
\mathcal{P}^x_\Delta\,. \label{pconst}
\end{eqnarray}
Constant moment maps fulfilling the above relation appear in the action
as coefficients of Fayet--Iliopoulos terms. Non trivial solutions of eq.
(\ref{pconst})
 exist only in correspondence to the following simple factors inside $\mathcal{G}$:
\begin{eqnarray}{\rm U}(1) &\leftrightarrow & \mathcal{P}^x_\Delta = g\,\delta^{x 3}\,%\delta_{\Delta 3}
\,\,\,\,\,\,\,;\,\,\,\,\,\,\,\,\,{\rm SO}(3)\leftrightarrow   \mathcal{P}^x_\Delta = g\,\delta^x_\Delta\,.\end{eqnarray}
The scalar potential is bilinear in the fermion shifts and the only
negative contribution is related to the gravitino:
\begin{eqnarray}
V(\phi) &=& \left(g_{\alpha\bar{\beta}}k^\alpha_\Lambda k^{\bar{\beta}}_\Sigma+4h_{uv}k^u_\Lambda k^{v}_\Sigma\right)\bar{L}^\Lambda L^\Sigma+\left(U^{\Lambda\Sigma}-3\bar{L}^\Lambda L^\Sigma\right)\mathcal{P}^x_\Lambda\mathcal{P}^x_\Sigma\,,\nonumber\\
L^\Sigma (z,\bar{z})& =& e^{\mathcal{K}/2}X^\Sigma
(z)\,\,;\,\,\,\nabla_{\bar{\alpha}} L^\Sigma\equiv
\left(\partial_{\bar{\alpha}}-\frac12(\partial_{\bar{\alpha}}\mathcal{K})\right)L^\Sigma
=0\,.
\end{eqnarray}
\section{$\mathcal{N}=2$ models with stable dS vacua}

Let us discuss now the minimal ingredients which were found in \cite{our}
to be necessary for the construction of an $\mathcal{N}=2$ model with a
stable dS vacuum, restricting for simplicity to the case $r=0$ in which
no hyper-multiplets are present.
\begin{itemize}
\item{Choice of a Special K{\"a}hler manifold of the form:
\begin{eqnarray}
\mathcal{M}_{SK}&=&\frac{{\rm SU}(1,1)}{{\rm U}(1)}\times\frac{{\rm
SO}(2,n-1)}{{\rm SO}(2)\times {\rm SO}(n-1)}
\end{eqnarray}
to parametrize using the Calabi--Vesentini coordinates
$\{z^\alpha\}=\{S,\,y^a\}$ ($a=0,\dots, n-2$) in terms of which the upper
and lower components of the symplectic section have the form:
\begin{eqnarray}
X^\Lambda(z^\alpha)&=&\left(\matrix{\frac{1}{2}(1+y^2)\cr \frac{{\rm
i}}{2}(1-y^2)\cr y^a }\right)\,;\,\,\,\,F_\Lambda(z^\alpha)=S \,
\eta_{\Lambda\Sigma}\,X^\Sigma\,;\nonumber\\
\eta_{\Lambda\Sigma}&=&{\rm diag}(+,+,\stackrel{n-1}{\overbrace{-,\dots,-
}})\, \,\,;\,\,\,\,y^2=(y^0)^2+\dots +(y^{n-2})^2\,.
\end{eqnarray}}
\item{The presence in the gauge group $\mathcal{G}\subset\mathbb{G}={\rm SU}(1,1)\times {\rm SO}(2,n-1)$ of two simple factors $\mathcal{G}_0\times \mathcal{G}_1$  of which $\mathcal{G}_0$, with coupling constant $e_0$, has to be {\it non--compact} and {\it non--abelian} and $ \mathcal{G}_1$, with coupling constant $e_1$, has to admit a Fayet--Iliopoulos term. The only choice for $\mathcal{G}_0$ is ${\rm SO}(2,1)$ while $ \mathcal{G}_1$ can be either ${\rm U}(1)$ or ${\rm SO}(3)$. The F. I. term corresponding to $ \mathcal{G}_1$ has then to be added to the Lagrangian}
\item{Generalize the choice of the symplectic section by rotating the symplectic embedding of $\mathcal{G}_1$ with respect to $\mathcal{G}_0$. This is done
by means of a rotation ${\bf R}(\theta)$ which acts on the symplectic section $V(z)$ and has the  following properties:
\begin{eqnarray}
{\bf R}(\theta) \in {\rm Sp}(2n +2, \mathbb{R})/\mathbb{G}\,;\qquad {\bf
R}(\alpha)^{-1}\, \mathcal{G}\,{\bf R}(\theta)=\mathcal{G}\,.
\end{eqnarray}
Since by construction ${\bf R}(\theta)$ is not in $\mathbb{G}$, its
effect is to map theories that are equivalent at the {\em ungauged}
level. A similar rotation was introduced by de Roo and Wagemans in the
context of $\mathcal{N}=4$ supergravity \cite{deRoo:Wagemans} to
generalize the construction of models with (unstable) de Sitter and
anti--de Sitter vacua.}
\end{itemize}
In the sequel we shall analyse three models in which the above
prescriptions are adopted and comment on the structure of their vacua.
\subsection{Model n.1: three vector multiplets}
To start with, we shall consider the simplest model, which describes
supergravity coupled to $n=3$ vector multiplets and no hyper-multiplet.
There are four vector fields $A_\mu^\Lambda $, three of them from the
vector multiplets plus the graviphoton in the gravity multiplet. The
Calabi--Vesentini coordinates are denoted by $\{S,\,y^0,\,y^1\}$. The
global symmetry group $\mathbb{G}$, which is the isometry group of the
scalar manifold, is ${\rm SU}(1,1)\times {\rm SO}(2,2)$. The embedding of
$\mathbb{G}$ in ${\rm Sp}(8,\mathbb{R})$, which defines its action as an
electro--magnetic duality, is fixed by the choice of the
Calabi--Vesentini symplectic frame. In particular, only the ${\rm
SO}(2,2)$ factor in $\mathbb{G}$ is a symmetry of the action, since as a
duality transformation group it does not change
 electric charges into magnetic ones.
The gauge group therefore has to be chosen within ${\rm SO}(2,2)$ and
according to the prescriptions given above, the only possibility is
$\mathcal{G}={\rm SO}(2,1)\times {\rm U}(1)$. The vector fields transform
in the ${\bf 4}$ of ${\rm SO}(2,2)$ which decompose with respect to
$\mathcal{G}$ into the direct sum of the adjoint of ${\rm SO}(2,1)$ and
the adjoint of ${\rm U}(1)$. Accordingly, the ${\rm SO}(2,2)$ invariant
metric $\eta_{\Lambda\Sigma}$ splits into the Cartan--Killing forms of
${\rm SO}(2,1)$, ${\rm diag}(+,+,-)$, and of ${\rm U}(1)$, ${\rm
diag}(-)$. Next, we rotate the symplectic embedding of ${\rm U}(1)$ with
respect to ${\rm SO}(2,1)$ by acting on the symplectic section $V(z)$
with the following transformation:
\begin{eqnarray}
{\bf R}(\theta)&=&\left(\matrix{\bfone_3 & 0 & 0 & 0\cr 0 & \cos (\theta)
&0&\sin (\theta)\cr 0&0&\bfone_3 &0\cr 0& -\sin (\theta) & 0 &  \cos
(\theta)}\right)\in \frac{{\rm Sp}(8,\mathbb{R})}{{\rm SU}(1,1)\otimes
{\rm SO}(2,2)}\,.
\end{eqnarray}
The above matrix commutes with the gauge group and acts non trivially
only on the entry corresponding to the ${\rm U}(1)$ generator. We then
introduce the F.I. term for the  ${\rm U}(1)$ factor:
$\mathcal{P}_\Lambda^x = e_1\, \delta^{x 3}\,\delta_{\Lambda 3}$.\par
\begin{figure}
\begin{center}
\epsfxsize =10cm
%\vskip 10cm
{\epsffile{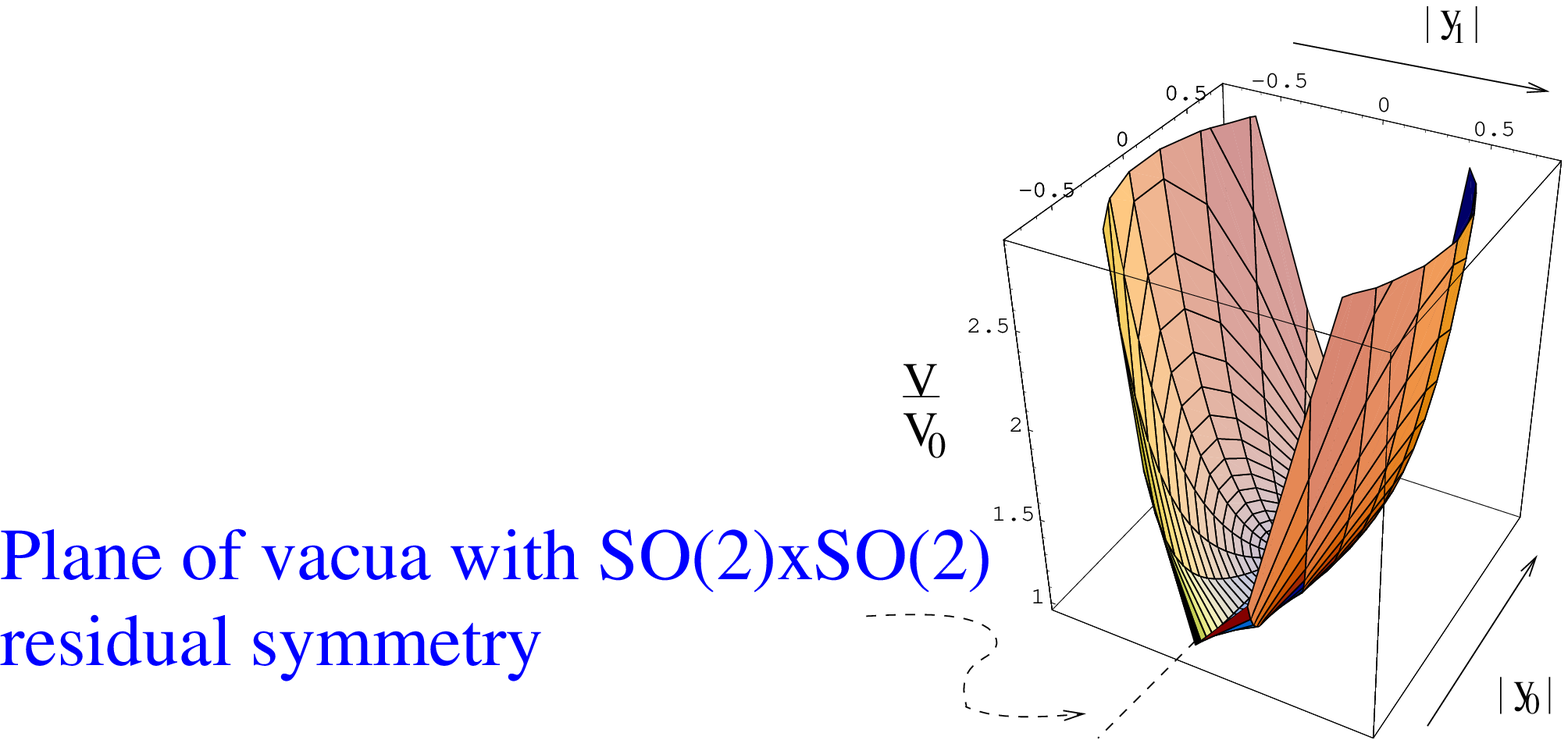}
%\vskip -8.5cm
}
% \hskip 2cm
%  \unitlength=1.1mm
\caption{\label{shape}}
\end{center}
\end{figure}

 The
resulting scalar potential turns out to be positive and $y^0$ to be a
flat direction. We also observe that $\partial_y V(S,y)_{|y^1=0}= 0$ and
the expression of the potential at $y^1=0$ is:
\begin{eqnarray}
V(S,y_0,y_1=0)&=& -\frac{e_0^2}{2 {\rm Im}(S)}-\frac{e_1^2}{2 {\rm
Im}(S)}\vert \cos (\theta)- S \sin (\theta)\vert^2 > 0\,.\label{pot}
\end{eqnarray}
It is positive since ${\rm Im}(S)<0$ is the positivity domain of the Lagrangian. It has an extremum at:
$$\phi^{(0)}\equiv\{S^{(0)}=\left(\frac{\cos (\theta)}{\sin (\theta)}-{\rm i}\vert\frac{e_0}{e_1\sin (\theta) } \vert\right),\,y^{0(0)},\,y^{1(0)}=0\}\,.$$
The second term in (\ref{pot}) contains  the F. I. term. We may now
appreciate the importance of the F. I. term and of the symplectic
rotation $\theta\neq 0$ in order for the potential to have a critical
point at finite $S$. The extremal value of $V$ is:
\begin{eqnarray}
V_0 &=& V(\phi^{(0)})=\vert e_0 e_1\sin (\theta)\vert > 0\,.
\end{eqnarray}
We have therefore found a plane of de Sitter critical points parametrized
by $\langle y^0\rangle$. In figure \ref{shape}, the potential is plotted
against $\vert y^a\vert$.
In these critical points there is a BEH mechanism at work according to
which the $y^0$ excitations are ``eaten'' by the vectors $A_\mu^0$ and
$A_\mu^1$ gauging the ${\rm SO}(2,1)$ non--compact generators, which then
acquire a mass $M^2=V_0/4$. There are thus no massless scalar excitations
in the effective theory at any of these critical points and the scalar
mass spectrum corresponds to the eigenvalues $k=3,6>0$ of the de Sitter
Casimir $C_2$. Therefore this model admits a plane of {\it stable} de
Sitter vacua with ${\rm SO}(2)\times {\rm SO}(2)$ residual symmetry.
\subsection{Models n.2 and 3: five vector multiplets without and with coupling to hyper--multiplets}
We have also considered a model with $n=5$ and $r=0$ in which the six
vector fields gauge the group $\mathcal{G}={\rm SO}(2,1)\times {\rm
SO}(3)$. According to our prescription the symplectic embedding of the
${\rm SO}(3)$ factor is rotated with respect to ${\rm SO}(2,1)$ and an
${\rm SO}(3)$ F.I. term is introduced. The resulting potential has
similar features as in model n.1, namely there is a plane parametrized by
$\langle y^0\rangle$ of stable de Sitter vacua, this time with residual
symmetry ${\rm SO}(2)\times {\rm SO}(3)$, in which the $y^0$ excitations
are Goldstone bosons for the non--compact gauge transformations. The
scalar mass spectrum is described by the same positive $C_2$
eigenvalues.\par In a third model we have introduced two
hyper--multiplets: $n=5$ and $r=2$. The quaternionic manifold is
$\mathcal{M}_{QK}=\frac{{\rm SO}(4,2)}{SO(4)\times {\rm SO}(2)}$ and the
gauge group is $\mathcal{G}={\rm SO}(2,1)\times {\rm SO}(3)\subset {\rm
Isom}(\mathcal{M}_{SK})\cap{\rm Isom}(\mathcal{M}_{QK}) $, that is we
choose $\mathcal{G}$ to have a diagonal action on the scalars in the
vector multiplets and in the hyper--multiplets. The same symplectic
rotation is performed as in model n.2. In this case we observe that on
the hypersurface $q^u=0$ $\partial_{q}V_{|q=0}=0$  and $V(S,y^a,q^u=0)$
has the same form as in model n.2. Moreover the moment maps at $q^u=0$
provide effective ${\rm SO}(3)$ F. I. terms which in model n.2 had to be
introduced by hand. As in model n.2 we find de Sitter vacua with residual
symmetry $SO(2)\times SO(3)$ in which a BEH mechanism involving a
combination of the excitations along $y^0$ and along a direction in the
quaternionic manifold is at work. Nevertheless there are still zero modes
in the quaternionic sector and the scalar mass spectrum corresponds to
the values $k=0,2,3,6$. Therefore, a definite statement about stability
of these vacua would require the computation of quantum corrections.\par
In all the three models considered above, supersymmetry is completely
broken on the de Sitter solution. This is a general feature of de Sitter
solutions in supergravity.
\section{Conclusions}
The models constructed above are the only instance of extended supergravities
with stable de Sitter vacua and which therefore can provide from the point of view of cosmology a description of the present stage of evolution of our
universe as driven towards a stable de Sitter regime \cite{kallosh:linde2}.
It is a challenging problem to lift these models to higher supersymmetry theories like $\mathcal{N}=4,\,8$ which have a closer relation to string/M--theory.
 In these latter theories, with respect to the $\mathcal{N}=2$ case, there is
an even more severe restriction on the possible gaugings imposed by supersymmetry. In
 particular although in the $\mathcal{N}=8$ theory new interesting gaugings have been recently found \cite{Andria:Dau:Fer:LLedo},  \cite{Hull}, \cite{deWit:I:Samtleben}, it is an open question whether symplectic rotations of the kind ${\bf R}(\theta)$ are allowed at all. A systematic analysis of this symplectic
freedom and of viable gaugings in maximal supergravities is given in
\cite{deWit:I:Samtleben}. Another interesting problem would be to
understand why in all the known examples of de Sitter solutions in
extended supergravities
 the scalar mass spectrum  appears to be quantized in units of the cosmological constant.


\section*{References}
\begin{thebibliography}{10}
\bibitem{experiment} Riess A G {\it et al.} 1998 {\it Astron.\ J.} {\bf 116}1009, Perlmutter S {\it et al.} 1999 {\it Astron.\ J.} {\bf 517} 565, Sievers J L  {\it et al.} 2002 {\it Preprint} astro-ph/0205387
\bibitem{kallosh:linde2} Kallosh R, Linde A, Prokushkin S, Shmakova M 2002  {\it Preprint} hep-th/0208156
\bibitem{linde90} Linde A D 1990 {\it Particle Physics and Inflationary Cosmology} (Switzerland: Harwood Academic)
\bibitem{dSexamples} de Wit B, Lauwers P G, Philippe R and Van Proeyen A 1984 {\it Phys.\ Lett.} B {\bf 135} 295, Castellani L, Ceresole A, Ferrara S, D'Auria R, Fr{\'e} P and Maina E 1986 {\it Nucl.\ Phys.} B {\bf 268} 317,  Gates S J and Zwiebach B 1983 {\it Phys. Lett} B {\bf 123} 200, Hull C M and Warner N P  1985 {\it Nucl. Phys.}  B {\bf 253} 675
\bibitem{deRoo:Wagemans} de Roo M and Wagemans P 1985
  {\it Nucl. Phys.} {\bf B262} 644
\bibitem{HulletAl} Hull C M and Warner N P 1988 {\it Class.\ Quantum\ Grav.}
{\bf 5} 1517,  Gibbons G W and Hull C M 2001 {\it Preprint} hep-th/0111072
\bibitem{kallosh} Kallosh R 2002 {\it Preprint} hep-th/0205315
\bibitem{kallosh:linde1} Kallosh R, Linde A, Prokushkin S, Shmakova M 2002 {\it Phys.\ Rev.\ D} {\bf 65} 105016
\bibitem{our} Fr{\'e} P, Trigiante M and Van Proeyen A 2002 {\it Class.\ Quantum\ Grav.} {\bf 19} 4167
\bibitem{deWit:Lau:VP} de Wit B, Lauwers P G and Van Proeyen A 1985
{\it Nucl. Phys.}{\bf B255} 569
\bibitem{n2gen}Andrianopoli L, Bertolini M, Ceresole A, D'Auria R, Ferrara S,
Fre P and Magri T 1997 {\it J.\ Geom.\ Phys.} {\bf 23} 111
\bibitem{Andria:Dau:Fer:LLedo} Andrianopoli L, D'Auria R, Ferrara F and Lledo M A 2002 {\it JHEP} {\bf 0207} 010
\bibitem{Hull} Hull C M 2002 {\it Preprint} hep-th/0204156
\bibitem{deWit:I:Samtleben} de Wit B, Samtleben H and Trigiante M 2002 {\it Preprint} hep-th/0212239 
\end{thebibliography}
\end{document}